# Beyond citations: Scholars' visibility on the social Web[1]

Judit Bar-Ilan*, Stefanie Haustein**, Isabella Peters***, Jason Priem****,
Hadas Shema* and Jens Terliesner***

**Judit.Bar-Ilan@.biu.ac.il; dassysh@gmail.com*
Department of Information Science, Bar-Ilan University, Ramat-Gan, 52900 (Israel)

** *s.haustein@fz-juelich.de*
Central Library, Forschungszentrum Jülich, Jülich, 52425 (Germany)

****isabella.peters@hhu.de; jens.terliesner@hhu.de*
Department of Information Science, Heinrich-Heine-University, Universitätsstr. 1, Düsseldorf, 40225 (Germany)

**** *priem@email.unc.edu*
School of Information & Library Science, University of North Carolina at Chapel Hill, 216 Lenoir Drive, CB #3360100 Manning Hall, Chapel Hill (USA)

**Abstract**
Traditionally, scholarly impact and visibility have been measured by counting publications and citations in the scholarly literature. However, increasingly scholars are also visible on the Web, establishing presences in a growing variety of social ecosystems. But how wide and established is this presence, and how do measures of social Web impact relate to their more traditional counterparts? To answer this, we sampled 57 presenters from the 2010 Leiden STI Conference, gathering publication and citations counts as well as data from the presenters' Web "footprints." We found Web presence widespread and diverse: 84% of scholars had homepages, 70% were on LinkedIn, 23% had public Google Scholar profiles, and 16% were on Twitter. For sampled scholars' publications, social reference manager bookmarks were compared to Scopus and Web of Science citations; we found that Mendeley covers more than 80% of sampled articles, and that Mendeley bookmarks are significantly correlated (r=.45) to Scopus citation counts.

## Introduction

Traditionally, the measurement of authors' impact or visibility is based on counting how often a particular author can be found in the reference lists of scientific publications. Information on the number of publications and citations of an author are published in citation databases such as Web of Science (WoS) or Scopus revealing how well an author is perceived in the scientific community. These reference-based databases differ in the number of analyzed publication sources with WoS having a smaller number of sources than Scopus (Moed & Visser, 2008). Citation counts are established in author evaluation but they reflect only half the truth: they just capture the author's impact on other authors, i.e. people who also publish scientific texts in particular sources. The author's impact on non-authors, i.e. pure readers, is missed in traditional citation counts (Haustein, 2012). Moreover, authors create "footprints" via profiles in social networks, homepages, or publication lists to make themselves and their work more visible. These aspects of scholarly behavior are also not acknowledged by citation counts. With social media, new platforms come into play which provide alternatives to gain a holistic view on the visibility and impact of authors. Such alternative approaches are summarized under the concept "altmetrics", where altmetrics is "the creation and study of new metrics based on the Social Web for analyzing, and informing scholarship" (Priem, Taraborelli, Groth, & Neylon, 2010). The basic concepts and the platforms used in this study as well as the related literature and the research questions guiding our study will be explained in the following sections.

[1] Judit Bar-Ilan and Hadas Shema were supported in this study by the FP-7 EU funded project ACUMEN on assessing Web indicators on research evaluation.

*Altmetrics and Alternative Platforms*

"Altmetrics" is used as an umbrella term which condenses ideas on how to combine social media with aspects of traditional scholarly practice (Priem et al., 2010); as such it is properly a subset of webometrics (Almind & Ingwersen, 1997; Thelwall, Vaughan, & Björneborn, 2005), although using new data sources and methods. The aim of altmetrics is to expand our views on impact, by considering new data sources and metrics. Altmetrics considers all stages and products of scholarly research (from "social" literature search via Facebook to discussion of published results with readers via Twitter), including any impact a publication or an author may have on other people, e.g., retweeting a tweet, downloading or bookmarking an article, sharing a blog post in social networks, or following the author—in short, a scholars "scientific 'street cred'" (Cronin, 2001). As such, the altmetrics approach offers new ways to measure impact of authors and publications which may complement rather than replace traditional indicators for research evaluation. The combination of traditional and alternative metrics for research evaluation will provide more complete author or article profiles as it captures more dimensions of scientific practice.

Two particularly useful sources for altmetrics are Mendeley (mendeley.com) and CiteULike (citeulike.org). These are scholarly social bookmarking systems (Henning & Reichelt, 2008; Hull, Pettifer, & Kell, 2008; Reher & Haustein, 2010), where users can save references or share them in groups and describe references with freely chosen keywords, so-called tags. They can also search for literature via similar users, tags or references and can export their libraries in various formats (e.g., BibTex). According to self-reported numbers, Mendeley is much larger than CiteULike (CuL): 5.9 million papers in CuL vs. more than 34 million in Mendeley as of the beginning of March 2012.

Both social bookmarking systems use a bag model for resources, meaning that a particular resource can be simultaneously saved or bookmarked by several users. This functionality allows for counting resource-specific bookmarking actions like how many users saved a particular resource. CiteULike also lets us count number of tagging actions. The tag frequency distribution shows in which way readers perceive the bookmarked resource because tags mirror how readers understand the resource's content and also often what motivates bookmarking (Kipp, 2006). Moreover, tag frequency distributions reveal which tags are popular, allowing the "wisdom of crowds" (Surowiecki, 2004) to describe resources and perhaps evaluate their relevance (Jiang, He, & Ni, 2011; Peters, 2009). Tags can be aggregated at multiple levels, allowing for multidimensional journal (Haustein, 2012) and author evaluation.

Li, Thelwall, and Giuistini (2012) investigated how bookmarks in Mendeley and CuL reflect papers' scholarly impact; they search Mendeley, CuL, and Web of Science (WoS) for 1,613 Nature and Science articles published in 2007. They found that 92% of sampled articles had been bookmarked by at least one Mendeley user, and 60% by one or more CuL users. Mendeley bookmarks showed moderate correlation to WoS (r=.55) and Google Scholar (r=.60) citations, with CuL correlations somewhat lower (WoS: .34, GS: .39), perhaps due to sparser data. The authors concluded that social bookmarking systems are valuable sources for measuring research impact from the readers' point of view.

Bar-Ilan (2011) studied the items tagged with "bibliometrics" on Mendeley and CuL and found that the author with the largest number of publications tagged with "bibliometrics" on both reference managers was quite unknown within the bibliometric community. In terms of publication sources the findings were much less surprising: the most frequently tagged source was *Scientometrics*.

Haustein, Golov, Luckanus, Reher, and Terliesner (2010) introduced the analysis of usage data from social bookmarking platforms as an alternative to measure journal perception. They focused on the evaluation of journal readership against the background of global download statistics being inaccessible (Gorraiz & Gumpenberger, 2010; Schlögl & Gorraiz, 2010). Bookmarks of journal articles on the STM-specialized platforms CuL, Connotea and BibSonomy were used to evaluate usage of 45 physics journals. Tags assigned to bookmarked publications were cumulated on journal level to reflect a reader-specific view on journal content.

Investigators have also examined other sources for altmetrics. Twitter has been the subject of several informetric and scientific communication related studies. Priem and Costello (2010) and Priem, Costello, and Dzuba (2011) found that scholars use Twitter as professional medium for sharing and discussing articles, while Eysenbach (2011) showed that highly-tweeted articles were 11 times more likely to later be highly-cited. Weller and Puschmann (2011), and Letierce, Passant, Decker, and Breslin (2010) analyzed the use of Twitter during scientific conferences and revealed that there was discipline-specific tweeting behavior regarding topic and number of tweets as well as references to different document types (i.e., blogs, journal articles, presentation slides). Other studies have examined altmetrics drawn from citations from Wikipedia articles (Nielsen, 2007) and blogs (Groth & Gurney, 2010, Shema & Bar-Ilan, 2011).

*Research Questions*

The following research questions are guiding our study:

1) What basic information on the researchers can be found on the Web? Do they have homepages, complete lists of publications, Google Citations profiles, and do they actively participate in social media platforms, such as Twitter and LinkedIn?

2) How does the altmetrics data compare with traditional metrics like publication and citation counts? For the comparison we focus on Mendeley and CuL data in the current study. What is the coverage of these reference managers, and what is the correlation between the number of readers and the number of citations at the article level?

3) What additional information can we learn from the available altmetrics data? Can user-assigned keywords, i.e. tags, expose qualitative information on authors?

## Data collection for the STI 2010 presenters

In this paper we considered the presenters at the STI 2010 conference that took place in Leiden. The names of the presenters appear in bold in the conference program (http://www.cwts.nl/pdf/STI_2010_Conference_Programme_0818.pdf). Ambiguous names of presenters, and presenters on whom no information was found in WoS and Scopus were excluded (altogether 10 names were excluded, e.g. Sabir Hicham). Herbert Marsh was also excluded, because his focus is not on topics related to bibliometrics and scientometrics. Thus our dataset consists of 57 researchers.

We manually collected basic Web-based information about the presenters, including the researchers' homepages (institutional and/or personal), their lists of publications published by them or by their institution, the existence of LinkedIn (linkedin.com) and Google Scholar Citations (scholar.google.com/citations) profiles, Twitter (twitter.com) account and Mendeley membership. Results were verified according to self-provided details such as name, profession, institute, publications, country and photo. Indecisive and ambiguous information was not included.

The publications of these researchers were searched on WoS and Scopus. All the searches were carried out during the second half of February 2012. On WoS we often had to limit the searches to the Information Science and Library Science categories in order to disambiguate several authors with identical names and initials. Scopus allows searching with the first name as well, which allowed us to identify the authors more easily. Thus we used the Scopus data as our main source for publication information. In some of the searches the DOIs of the articles were needed. There were a large number of DOIs missing in Scopus; these were supplemented with data from WoS, and also from Crossref (crossref.org/guestquery). The searches were limited to articles, articles in press, conference papers and reviews on Scopus, and to articles, reviews, notes and proceedings papers on WoS. We identified 1,574 publications (including doubles, i.e., two or more presenters coauthored the same publication) of the 57 authors, 1,351 of which had a DOI. 1,136 of these publications were unique, i.e., 215 documents were co-publications. Citation data from Google Scholar are not presented, because of the need for extensive data cleansing and identification of the specific type of publications. Microsoft Academic Search was not utilized, because at this point of time it does not offer comprehensive information.

Mendeley publication and readership information was retrieved manually from mendeley.com. We gathered data from ReaderMeter (readermeter.org) and total-impact (total-impact.org), two search tools built on the Mendeley API. ReaderMeter takes an author's name and returns author-level information extracted from Mendeley; total-impact reports usage in Mendeley, CuL, Wikipedia, Twitter, and many other social media on a per-article basis, using DOI, PubMed ID, or other identifiers. We also gathered Mendeley data by hand using the Mendeley Web search interface. This latter approach proved much more comprehensive, because we were able to search by title and author rather than just DOI, surfacing additional entries; in all, 33% of sample articles bookmarked in Mendeley were without DOI. We also found many duplicated entries that were not returned by the API.

In CuL, publications can be searched by DOI. As DOIs allow for easy identification of articles, bookmarks for the 1,136 documents were retrieved via the DOI. However it should be noted that the bibliographic data in CuL or Mendeley, and as mentioned above also in Scopus and WoS, is incomplete. Haustein and Siebenlist (2011) found that 77.6% percent of bookmarks to physics papers in CuL contained the correct DOI. The number of articles bookmarked in CuL might thus be higher as from what was retrieved via DOI.

This detailed description of the data collection process emphasizes its limitations, but also shows that efforts were made to improve the quality of the dataset.

## Results

The following section summarizes the results of our study. We will give an overview on the presenters' footprints made by themselves, i.e. self-initiated author information, followed by the information on traditional and alternative author metrics in terms of authors' article visibility. Additionally, we will compare article visibility in traditional citation-based databases and social bookmarking systems.

### *Author level metrics and footprints*

In this section we present findings based on searches for author names. We searched traditional databases (WoS and Scopus) and self-initiated author information found on the Web and social media. Loet Leydesdorff and Wolfgang Glänzel achieved the highest citation and publication counts and h-indexes. We defined a 'homepage' as a page which is hosted either by the researcher's workplace or is part of the researcher's own domain and has at least a short CV. Forty-eight of the researchers in our sample (84%) have at least one institutional homepage, 6 (11%) have none, and 3 (5%) share a page. Six researchers (11%) also have personal homepages. A narrow majority of researchers (53%) had full lists of publications on either their institutional or professional pages, with another 33% displaying only a selected list. Only 14% (8 researchers) had no publication lists at all. Author level metrics and web-based information created by the presenters or their institution appear in Table 1.

Google Scholar (GS) launched a new service called "Google Scholar Citations" on November 16, 2011, which allows researchers to create an editable, verified (using an institutional email) profile including their personal details, a list of their papers, and citations to those papers. One can choose whether to make the profile public or private (Connor, 2011). Thirteen of the 57 researchers in the sample (23%) have public GS profiles and additional researchers in our set may have private profiles. Even though the GS profiles are fairly new, almost a quarter of the researchers chose to verify their profile and make it public, which could be an indication of the service's future potential. LinkedIn seems to be a very popular social media service, with 40 presenters (70%) having a LinkedIn account. Additionally, nine of the researchers (16%) have Twitter accounts. While lower than the percentages on LinkedIn and GS, this percentage is higher than the conservative estimate of 2.5% of scholars on Twitter found by Priem, Costello, and Dzuba (2011) in a sample drawn from five representative UK and US universities; this may be partly due to this smaller sample allowing more thorough searches for Twitter profiles. In Table 1 "*y*" represents that the researcher has a presence on the specific platform.

### *Article level metrics – traditional and alternative*

Another aspect of author visibility is based on the presenters' articles and their coverage in traditional and social databases. We use the subset of articles with DOIs found in Scopus as basis for comparisons of traditional and alternative metrics. In CuL and Mendeley users can bookmark articles for themselves, but they may also save the article for a group. In our study bookmarks by groups are not considered so that each bookmark reflects usage by one user. Mendeley and CuL also retrieve articles which were not yet bookmarked by any readers. In Mendeley these entries are created by the authors of the paper and in CuL these seem to be generated by the database. These entries without users were excluded as well.

Table 1. Author level information and metrics.

| Name | # publications in WoS | # citations in WoS | H-index in WoS | Scopus# publications in | # citations in Scopus | H-index in Scopus | Homepage (personal (p)); institutional (i); group page only (g) | List of publications: seemingly full (f), selected/partial (s), or not found (n) | GoogleCitations profile | LinkedIn account | Twitter account |
|---|---|---|---|---|---|---|---|---|---|---|---|
| Aguillo, Isidro | 32 | 143 | 7 | 41 | 196 | 7 | i | s | y | y | n |
| Aksnes, Dag W | 13 | 295 | 8 | 14 | 325 | 8 | i | f | n | y | n |
| Andersen, Jens Peter | 2 | 1 | 1 | 4 | 0 | 0 | p+i | f | n | y | y |
| Archambault, Éric | 15 | 185 | 8 | 15 | 201 | 9 | i | f | y | y | n |
| Bar-Ilan, Judit | 68 | 871 | 19 | 72 | 1092 | 20 | i | f | y | y | y |
| Bonaccorsi, Andrea | 25 | 549 | 13 | 36 | 564 | 13 | i | f | n | y | n |
| Bornmann, Lutz | 65 | 849 | 15 | 84 | 975 | 16 | p+i | f | n | y | n |
| Boyack, Kevin | 23 | 569 | 9 | 28 | 689 | 11 | g | s | n | y | n |
| Bubela, Tania | 12 | 196 | 8 | 16 | 142 | 6 | i | s | n | y | n |
| Buter, Renald | 8 | 66 | 4 | 9 | 69 | 4 | i | s | n | y | n |
| Butler, Linda | 14 | 239 | 9 | 23 | 346 | 11 | i | s | n | n | n |
| Daniel, Hans-Dieter | 44 | 762 | 14 | 62 | 973 | 17 | i | f | n | n | n |
| De Filippo, Daniela | 4 | 6 | 2 | 7 | 11 | 2 | i | f | n | n | n |
| Engels, Tim | 3 | 7 | 2 | 4 | 9 | 2 | i | f | y | n | n |
| Foray, Dominique | 19 | 182 | 8 | 20 | 296 | 10 | i | f | n | y | n |
| Gerritsma, Wouter | 3 | 20 | 2 | 4 | 24 | 2 | i | f | y | y | y |
| Gingras, Yves | 46 | 300 | 9 | 39 | 379 | 11 | i | s | n | y | n |
| Glänzel, Wolfgang | 139 | **3305** | **31** | 143 | 2825 | 28 | i | f | n | y | n |
| Gorraiz, Juan | 17 | 36 | 4 | 23 | 35 | 3 | p+i | s | n | y | n |
| Gumpenberger, Christian | 3 | 0 | 0 | 6 | 2 | 1 | not found | f | n | y | n |
| Hardeman, Sjoerd | 1 | 18 | 1 | 1 | 17 | 1 | i | n | n | y | n |
| Harnad, Stevan | 44 | 1164 | 11 | 65 | 1371 | 17 | i | f | y | y | y |
| Haustein, Stefanie | 3 | 1 | 1 | 3 | 3 | 1 | i | f | n | y | y |
| Hoekman, Jarno | 2 | 25 | 2 | 6 | 44 | 4 | i | f | n | y | n |
| Klavans, Richard | 16 | 313 | 8 | 14 | 314 | 8 | g | s | n | y | n |
| Kousha, Kayvan | 14 | 156 | 6 | 14 | 165 | 6 | p+i | f | y | y | n |
| Kuan, Chung-Huei | 2 | 2 | 1 | 4 | 2 | 1 | not found | n | n | n | n |
| Larivière, Vincent | 24 | 212 | 9 | 24 | 238 | 11 | i | f | y | n | n |
| Leten, Bart | 2 | 12 | 2 | 2 | 13 | 2 | i | s | n | n | n |
| Levitt, Jonathan M | 8 | 34 | 3 | 10 | 35 | 3 | i | s | n | n | n |
| Leydesdorff, Loet | **162** | 2885 | 28 | **176** | **3420** | **29** | p+i | f | y | y | n |
| Luwel, Marc | 17 | 255 | 11 | 19 | 316 | 12 | not found | n | n | y | n |
| Marchant, Thierry | 22 | 92 | 6 | 28 | 117 | 6 | i | f | n | n | n |
| Mauleón, Elva (Elba) | 3 | 23 | 2 | 3 | 26 | 2 | i | n | n | n | n |
| Neufeld, Jörg | 2 | 8 | 2 | 2 | 8 | 2 | i | f | n | n | n |
| Noyons, Ed | 21 | 338 | 11 | 24 | 311 | 10 | i | f | n | y | y |
| Paier, Manfred | 1 | 1 | 1 | 1 | 1 | 1 | not found | n | n | y | y |
| Picard-Aitken, Michelle | 2 | 17 | 2 | 3 | 21 | 2 | not found | n | n | y | n |
| Porter, Alan L. | 70 | 556 | 12 | 92 | 783 | 16 | i | s | n | n | n |
| Rafols, Ismael | 22 | 477 | 12 | 22 | 481 | 12 | i | s | y | y | n |
| Sandstrom, Ulf | 6 | 54 | 3 | 8 | 78 | 4 | i | f | y | y | n |
| Schloegl, Christian | 16 | 66 | 5 | 13 | 28 | 3 | i | f | n | n | n |
| Schmoch, Ulrich | 23 | 333 | 10 | 36 | 565 | 11 | i | s | n | y | n |
| Schneider, Jesper W | 9 | 51 | 4 | 10 | 55 | 5 | i | s | n | y | n |
| Schubert, Torben | 9 | 47 | 4 | 14 | 63 | 4 | i | f | n | n | y |
| Shelton, Robert D. | 3 | 36 | 2 | 4 | 43 | 2 | p+i | s | n | n | n |
| Small, Henry | 41 | 2624 | 21 | 33 | 1240 | 16 | g | n | n | y | n |
| Tunger, Dirk | 4 | 8 | 2 | 7 | 28 | 3 | i | n | n | n | n |
| Van Eck, Nees Jan | 24 | 151 | 8 | 30 | 190 | 9 | p+i | f | y | y | y |
| Van Leeuwen, Thed | 41 | 984 | 18 | 52 | 1011 | 19 | i | f | n | y | n |
| Van Looy, Bart | 22 | 324 | 10 | 24 | 378 | 11 | i | f | n | y | n |
| Van Raan, Anthony | 88 | 2557 | 30 | 97 | 2010 | 27 | i | s | n | y | n |
| Van Vught, Frans A | 5 | 34 | 4 | 9 | 58 | 5 | i | s | n | y | n |
| Waltman, Ludo | 22 | 155 | 8 | 28 | 178 | 8 | p+i | f | y | y | n |
| Yegros-Yegros, Alfredo | 2 | 13 | 2 | 4 | 14 | 2 | i | s | n | y | n |
| Zitt, Michel | 25 | 443 | 14 | 28 | 432 | 13 | not found | s | n | n | n |
| Zuccala, Alesia | 13 | 80 | 4 | 14 | 88 | 6 | i | f | n | y | n |

Table 2A. Summary of metric coverage. "Events" are either bookmarks or citations, depending on the database.

| | Number of indexed documents | Total event counts | Percent sampled articles with nonzero event counts (total) | Mean events per article with nonzero count (sd) |
|---|---|---|---|---|
| Scopus | 1,136 | 18,755 | **85%** (961) | 19.5 (39.0) |
| Mendeley | 928 | 8,847 | **82%** (928) | 9.5 (13.4) |
| WoS | 957 | 17,858 | **74%** (845) | 21.1 (46.5) |
| CiteULike | 319 | 777 | **28%** (319) | 2.4 (2.7) |

Table 2A and 2B present the results found for the subset of 1,136 articles. The coverage of these documents by Mendeley was quite good: 928 (82%) of the documents had at least one Mendeley bookmark, similar to the 961 cited items (85%) in Scopus, and greater than the 845 (74%) articles with citations in WoS. By contrast, only 319 (28%) of articles were bookmarked in CuL; although coverage may be underestimated because bookmarks without a correct DOI were not retrieved, this does suggest Mendeley is cementing dominance in usage. This is not only reflected in the coverage of documents but also by the average activity on bookmarked documents: in Mendeley each bookmarked document was bookmarked by a mean of 9.5 users, compared to 2.4 bookmarkes in CuL. Unsurprisingly given Mendeley's very recent founding, older articles are less bookmarked. Of the 85 sample articles published before 1990, only 44% are bookmarked in Mendeley, while 88% of those published since 2000 have Mendeley bookmarks.

Table 2B. Author level alt-metrics.

| **Name** | **DOI# publications searched via** | **# citations in Scopus** | **citation rate in Scopus** | **# publications Mendeley** | **Mendeley coverage** | **# of bookmarks in Mendeley** | **bookmarks per publication in Mendeley** | **# publications CiteULike** | **CiteULike coverage** | **# of bookmarks in CiteULike** | **bookmarks per publication in CiteULike** | **# unique users in CiteULike** |
|---|---|---|---|---|---|---|---|---|---|---|---|---|
| Noyons, Ed | 23 | 307 | 13.3 | 23 | **100.0%** | 206 | 9.0 | 12 | 52.2% | 22 | 1.8 | 19 |
| Aksnes, Dag W | 14 | 325 | 23.2 | 14 | **100.0%** | 127 | 9.1 | 5 | 35.7% | 8 | 1.6 | 8 |
| Klavans, Richard | 14 | 314 | 22.4 | 14 | **100.0%** | 270 | 19.3 | 9 | 64.3% | 26 | 2.9 | 22 |
| Kousha, Kayvan | 13 | 165 | 12.7 | 13 | **100.0%** | 184 | 14.2 | 9 | 69.2% | 26 | 2.9 | 22 |
| Schneider, Jesper W | 10 | 55 | 5.5 | 10 | **100.0%** | 65 | 6.5 | 7 | 70.0% | 16 | 2.3 | 11 |
| Buter, Renald | 7 | 38 | 5.4 | 7 | **100.0%** | 57 | 8.1 | 5 | 71.4% | 10 | 2.0 | 10 |
| Sandstrom, Ulf | 6 | 78 | 13.0 | 6 | **100.0%** | 36 | 6.0 | 3 | 50.0% | 5 | 1.7 | 4 |
| De Filippo, Daniela | 5 | 6 | 1.2 | 5 | **100.0%** | 8 | 1.6 | 1 | 20.0% | 1 | 1.0 | 1 |
| Shelton, Robert D. | 4 | 43 | 10.8 | 4 | **100.0%** | 24 | 6.0 | 1 | 25.0% | 1 | 1.0 | 1 |
| Yegros-Yegros, Alfredo | 4 | 14 | 3.5 | 4 | **100.0%** | 23 | 5.8 | 1 | 25.0% | 1 | 1.0 | 1 |
| Mauleón, Elva (Elba) | 3 | 26 | 8.7 | 3 | **100.0%** | 13 | 4.3 | 0 | 0.0% | 0 | | 0 |
| Andersen, Jens Peter | 2 | 0 | 0.0 | 2 | **100.0%** | 8 | 4.0 | 1 | 50.0% | 1 | 1.0 | 1 |
| Gerritsma, Wouter | 2 | 19 | 9.5 | 2 | **100.0%** | 20 | 10.0 | 0 | 0.0% | 0 | | 0 |
| Haustein, Stefanie | 2 | 2 | 1.0 | 2 | **100.0%** | 35 | 17.5 | 1 | 50.0% | 4 | 4.0 | 4 |
| Leten, Bart | 2 | 13 | 6.5 | 2 | **100.0%** | 37 | 18.5 | 0 | 0.0% | 0 | | 0 |
| Picard-Aitken, Michelle | 2 | 21 | 10. | 2 | **100.0%** | 13 | 6.5 | 0 | 0.0% | 0 | | 0 |
| | | | 5 | | | | | | | | | |
| Schloegl, Christian | 2 | 9 | 4.5 | 2 | **100.0%** | 12 | 6.0 | 1 | 50.0% | 1 | 1.0 | 1 |
| Hardeman, Sjoerd | 1 | 17 | 17.0 | 1 | **100.0%** | 16 | 16.0 | 1 | **100.0%** | 4 | 4.0 | 4 |
| Paier, Manfred | 1 | 1 | 1.0 | 1 | **100.0%** | 3 | 3.0 | 0 | 0.0% | 0 | | 0 |
| Daniel, Hans-Dieter | 58 | 952 | 16.4 | 57 | 98.3% | 498 | 8.7 | 28 | 48.3% | 56 | 2.0 | 39 |
| Van Eck, Nees Jan | 29 | 189 | 6.5 | 28 | 96.6% | 185 | 6.6 | 12 | 41.4% | 20 | 1.7 | 11 |
| Van Leeuwen, Thed | 47 | 938 | 20.0 | 45 | 95.7% | 347 | 7.7 | 12 | 25.5% | 22 | 1.8 | 15 |
| Luwel, Marc | 16 | 244 | 15.3 | 15 | 93.8% | 102 | 6.8 | 5 | 31.3% | 6 | 1.2 | 4 |
| Waltman, Ludo | 28 | 178 | 6.4 | 26 | 92.9% | 199 | 7.7 | 11 | 39.3% | 17 | 1.5 | 9 |
| Boyack, Kevin | 26 | 688 | 26.5 | 24 | 92.3% | 538 | **22.4** | 14 | 53.8% | 44 | 3.1 | 35 |
| Bornmann, Lutz | 74 | 957 | 12.9 | 68 | 91.9% | 590 | 8.7 | 33 | 44.6% | 71 | 2.2 | 45 |
| Van Looy, Bart | 22 | 341 | 15.5 | 20 | 90.9% | 288 | 14.4 | 4 | 18.2% | 8 | 2.0 | 7 |
| Aguillo, Isidro | 32 | 179 | 5.6 | 29 | 90.6% | 203 | 7.0 | 9 | 28.1% | 21 | 2.3 | 13 |
| Bubela, Tania | 10 | 132 | 13.2 | 9 | 90.0% | 52 | 5.8 | 1 | 10.0% | 1 | 1.0 | 1 |
| Levitt, Jonathan M | 8 | 34 | 4.3 | 7 | 87.5% | 54 | 7.7 | 3 | 37.5% | 4 | 1.3 | 3 |
| Larivière, Vincent | 23 | 234 | 10.2 | 20 | 87.0% | 296 | 14.8 | 13 | 56.5% | 49 | 3.8 | 41 |
| Small, Henry | 30 | 1214 | **40.5** | 26 | 86.7% | 284 | 10.9 | 9 | 30.0% | 15 | 1.7 | 12 |
| Rafols, Ismael | 22 | 481 | 21.9 | 19 | 86.4% | 242 | 12.7 | 6 | 27.3% | 17 | 2.8 | 15 |
| Harnad, Stevan | 35 | 1229 | 35.1 | 30 | 85.7% | 636 | 21.2 | 13 | 37.1% | 81 | **6.2** | 65 |
| Archambault, Éric | 14 | 192 | 13.7 | 12 | 85.7% | 127 | 10.6 | 8 | 57.1% | 22 | 2.8 | 21 |
| Leydesdorff, Loet | **159** | **3044** | 19.1 | **134** | 84.3% | **1843** | 13.8 | **43** | 27.0% | **125** | 2.9 | **69** |
| Butler, Linda | 18 | 335 | 18.6 | 15 | 83.3% | 123 | 8.2 | 2 | 11.1% | 2 | 1.0 | 2 |
| Gingras, Yves | 34 | 372 | 10.9 | 28 | 82.4% | 401 | 14.3 | 15 | 44.1% | 72 | 4.8 | 59 |
| Zitt, Michel | 27 | 414 | 15.3 | 22 | 81.5% | 152 | 6.9 | 6 | 22.2% | 8 | 1.3 | 7 |
| Zuccala, Alesia | 10 | 70 | 7.0 | 8 | 80.0% | 67 | 8.4 | 3 | 30.0% | 6 | 2.0 | 6 |
| Bonaccorsi, Andrea | 34 | 558 | 16.4 | 27 | 79.4% | 329 | 12.2 | 4 | 11.8% | 9 | 2.3 | 9 |
| Bar-Ilan, Judit | 60 | 967 | 16.1 | 47 | 78.3% | 373 | 7.9 | 24 | 40.0% | 82 | 3.4 | 56 |
| Gumpenberger, Christian | 4 | 1 | 0.3 | 3 | 75.0% | 6 | 2.0 | 0 | 0.0% | 0 | | 0 |
| Tunger, Dirk | 4 | 13 | 3.3 | 3 | 75.0% | 19 | 6.3 | 1 | 25.0% | 3 | 3.0 | 3 |
| Porter, Alan L. | 59 | 761 | 12.9 | 44 | 74.6% | 386 | 8.8 | 12 | 20.3% | 24 | 2.0 | 19 |
| Glänzel, Wolfgang | 131 | 2700 | 20.6 | 97 | 74.0% | 601 | 6.2 | 34 | 26.0% | 69 | 2.0 | 45 |
| Schmoch, Ulrich | 30 | 561 | 18.7 | 22 | 73.3% | 166 | 7.5 | 2 | 6.7% | 2 | 1.0 | 2 |
| Foray, Dominique | 15 | 164 | 10.9 | 11 | 73.3% | 71 | 6.5 | 1 | 6.7% | 2 | 2.0 | 2 |
| Schubert, Torben | 13 | 62 | 4.8 | 9 | 69.2% | 55 | 6.1 | 3 | 23.1% | 4 | 1.3 | 4 |
| Van Raan, Anthony | 80 | 1620 | 20.3 | 55 | 68.8% | 420 | 7.6 | 18 | 22.5% | 39 | 2.2 | 27 |
| Hoekman, Jarno | 6 | 44 | 7.3 | 4 | 66.7% | 55 | 13.8 | 2 | 33.3% | 5 | 2.5 | 5 |
| Engels, Tim | 3 | 9 | 3.0 | 2 | 66.7% | 16 | 8.0 | 0 | 0.0% | 0 | | 0 |
| Van Vught, Frans A | 8 | 52 | 6.5 | 5 | 62.5% | 17 | 3.4 | 0 | 0.0% | 0 | | 0 |
| Marchant, Thierry | 26 | 116 | 4.5 | 15 | 57.7% | 30 | 2.0 | 1 | 3.8% | 2 | 2.0 | 2 |
| Gorraiz, Juan | 17 | 31 | 1.8 | 9 | 52.9% | 34 | 3.8 | 3 | 17.6% | 8 | 2.7 | 7 |
| Kuan, Chung-Huei | 4 | 2 | 0.5 | 2 | 50.0% | 15 | 7.5 | 1 | 25.0% | 1 | 1.0 | 1 |
| Neufeld, Jörg | 2 | 8 | 4.0 | 1 | 50.0% | 6 | 6.0 | 1 | 50.0% | 1 | 1.0 | 1 |
| **All publications** | **1136** | **18755** | **16.5** | **928** | **81.7%** | **8847** | **9.5** | **319** | **28.1%** | **777** | **2.4** | **291** |

In table 2B the number of bookmarks and the number of publications bookmarked in Mendeley and CiteULike are aggregated on the author level. The table is sorted descendingly by coverage in Mendeley and number of documents so that the presenter with the highest visibility in relation to his publication output is listed on top. For 19 authors all of their searched publications had at least one reader in Mendeley. With all of his 23 documents bookmarked in Mendeley, Ed Noyons can be regarded as the STI presenter with the highest relative visibility in Mendeley. Loet Leydesdorff reaches the highest values in all metrics based on absolute values, i.e. publications, citations and bookmarks and publications bookmarked in Mendeley and CuL. In contrast to Mendeley in CuL the number of unique users can be analyzed. Overall, 291 different users bookmarked 319 documents 777 times. Leydesdorff reached the largest audience; his publications were bookmarked by 69 different users. While the highest citation rate per paper was received by Henry Small, the greatest "altmetrics rates" are received by Kevin Boyack in Mendeley (22.4 bookmarks per publication) and Stevan Harnad in CuL (6.2 bookmarks per publication).

Contrary to traditional bibliographic and citation-based databases, social bookmarking systems like CuL, allow users to add their own keywords to bookmarked articles. These tags directly reflect how users perceive the content of the bookmarked article (Peters, Haustein & Terliesner, 2011). Aggregated by author, tags can show a unique, reader-centric view of authors' areas of expertise. To control for irregularities of spelling in (uncontrolled) tags, we unified spellings where appropriate; for example "open_access", "open-access" and "openaccess" were all unified under the third spelling variation. Examples for tag-based qualitative author information are displayed in Figures 1, and 2 for Loet Leydesdorf and Stevan Harnad, who are two of the authors with the highest number of tag applications. Because CuL applies a bag model for tags, these tag clouds show not only what the readers think each authors' work is about, but the term frequency (represented by font size) also indicates the perceived importance of different topics. The relative dominance of a single tag, "openaccess," in Harnad's cloud suggests that readers perceive his work as more focused compared to Leydesdorff, whose cloud suggests multiple strands of research including "triplehelix" and "collaboration" along with the more dominant "bibliometrics," "science," and "citation."

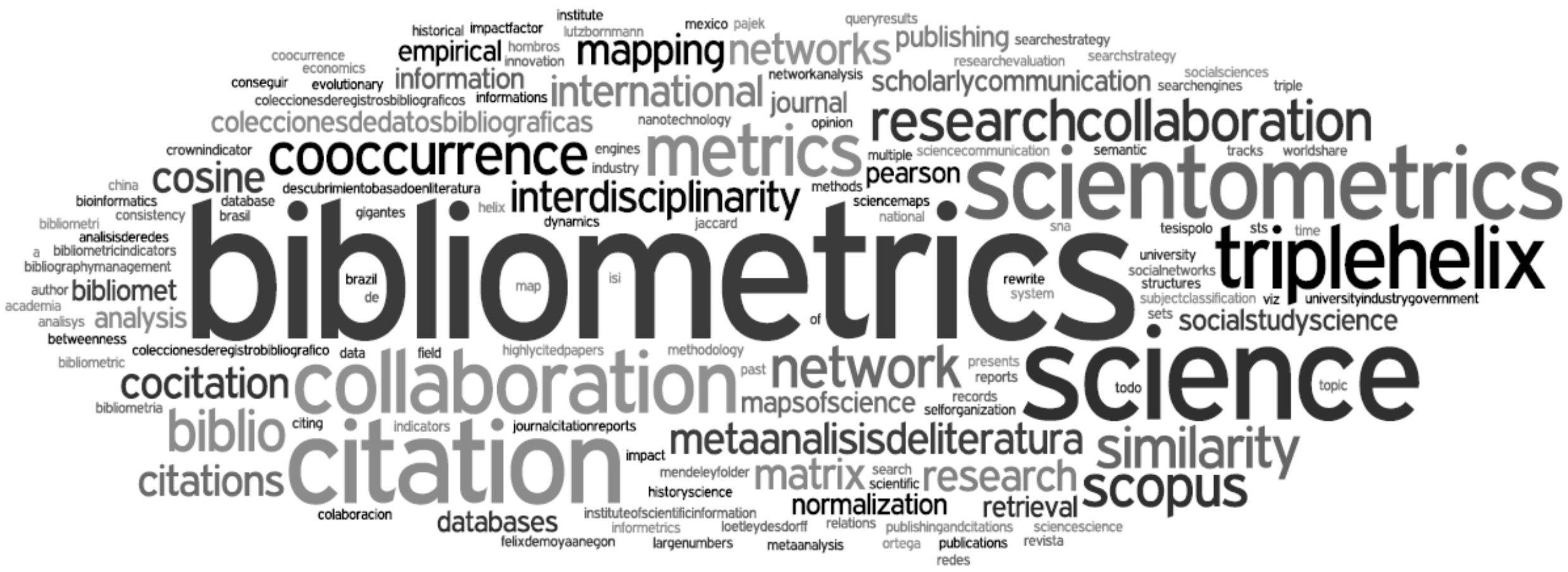

Figure 1: All tags assigned to articles published by Loet Leydesdorff.

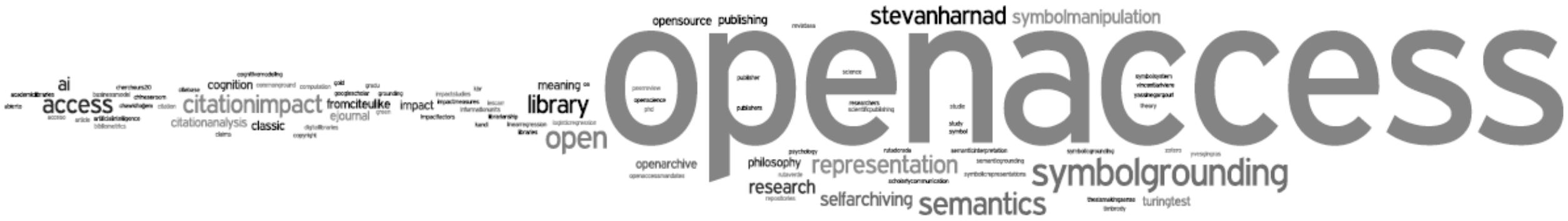

Figure 2: All tags assigned to articles published by Stevan Harnad.

*Comparison of traditional and alternative metrics in terms of author visibility*

A potential strength of altmetrics is that they track forms of impact not reflected in the citation record. Heavy bookmarking suggests that an article is being used or valued in some way; this impact is likely related to citation impact, but not identical. If this is true, it should be reflected in partial correlation between citations and bookmarks. This is in fact what we found in our sample articles, as shown in Table 3 and Figure 3.

Table 3. Correlations between citations and bookmarks for 1,136 documents.

| *Spearman's ρ* | citations (Scopus) | bookmarks (Mendeley) | bookmarks (CiteULike) |
|---|---|---|---|
| citations (Scopus) | | .448** | .232** |
| bookmarks (Mendeley) | .448** | | .441** |
| bookmarks (CiteULike) | .232** | .441** | |

*N=1136* ***. Correlation is significant at the 0.01 level (2-tailed).*

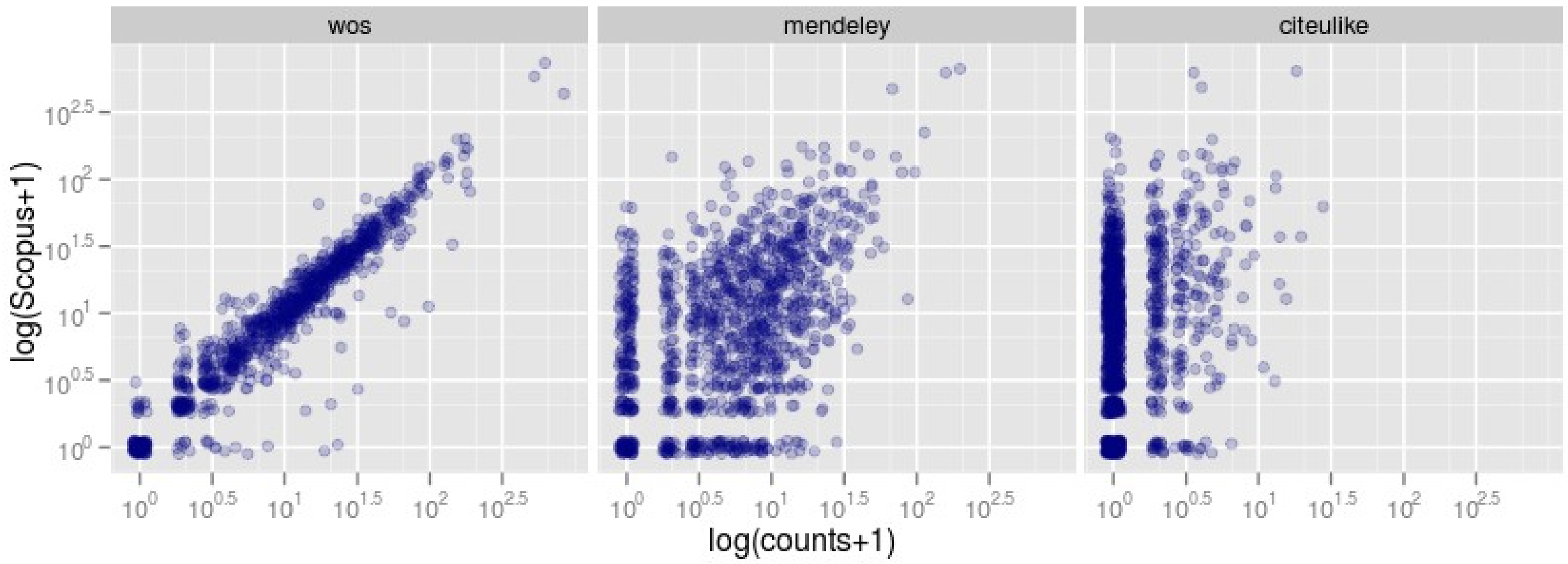

Figure 3: Relationships between log-normalized counts per article.

Both these are based on single articles to avoid misinterpretation of correlation values which may occur when working on author level. Table 3 shows Spearman correlations between Scopus, CuL, and Mendeley; the highest correlation is between Mendeley and Scopus at ρ=.448. This is slightly lower than Li et al. (2012) found for Nature and Science articles, but still of medium strength and significant. Figure 3 shows the relationships between log-normalized Scopus citation counts (y-axis) and log-normalized counts from the other three data sources. The plots show that as counts of citations and altmetrics grow, their relationship tightens, perhaps overwhelming sources of noise.

**Discussion & Future Work**

This study is limited by the specificity of its sample, as well as the relative messiness of social media sources; moreover, it will remain difficult to interpret altmetrics like Mendeley bookmarks until we have built a better understanding of how and why scholars use these tools. However, our findings are a useful and important early step toward understanding new online sources for impact data.

The researchers in our sample have a substantial professional web presence; 70% of them are on LinkedIn and 84% have institutional homepages. Though representing a professional presence, Google Scholar profiles (23%) were not as popular as LinkedIn accounts. However, in this case we can speculate that this relatively low percentage comes from its relatively short time in existence and from the need of both having Google account and an institutional email address for verification and retrieving by search engines. Twitter accounts were less common, with 16% of scholars having accounts on the service.

Users of online social reference managers have added substantial data about our sampled researchers' online impact. Mendeley covered 82% of our researchers' sampled outputs, and all 57 sampled authors had at least 50% of their publications bookmarked on Mendeley. With only 28.1% of documents covered, the readership in CuL was much lower, as also found by Li et al. (2012). However, CuL data on tags and tagging frequency offer a valuable, reader-centric perspective on authors' publication histories.

The total-impact tool provides much more article-level data including citation on blogs, Twitter, and Wikipedia which due to space constraints are not discussed in this paper, but will be analyzed in future work. Continued research in to these and other altmetric sources has substantial promise to help build a new "bibliometric spectroscopy" (Cronin, 2001), expanding and deepening our understanding of scholarly impact.